\documentclass[aps,pra,twocolumn,showpacs,amsmath,10pt]{revtex4}

\usepackage{bm}
\usepackage{graphicx}

\newcommand{\bra}[1]{\langle #1 | \,}
\newcommand{\ket}[1]{\, | #1 \rangle}

\newcommand{\expv}[1]{\langle #1 \rangle}

\newcommand{\la}{\lambda}

\newcommand{\de}{\delta}
\newcommand{\De}{\Delta}

\newcommand{\hlf}{\mbox{$\frac{1}{2}$}}

\newcommand{\ds}{\!\! \downarrow}
\newcommand{\Js}{\mathcal{J}}

\newcommand{\lra}{\leftrightarrow}
\newcommand{\hs}{\hat{\sigma}}
\newcommand{\ha}{\hat{a}}

\newcommand{\erf}{\operatorname{erf}}

\newcommand{\be}{\begin{equation}}
\newcommand{\ee}{\end{equation}}
\newcommand{\bea}{\begin{eqnarray}}
\newcommand{\eea}{\end{eqnarray}}
\newcommand{\beann}{\begin{eqnarray*}}
\newcommand{\eeann}{\end{eqnarray*}}
\newcommand{\besal}[1]{\begin{subequations}\label{#1}\begin{eqnarray}}
\newcommand{\besa}{\begin{subequations}\begin{eqnarray}}
\newcommand{\eesa}{\end{eqnarray}\end{subequations}}

\begin{document}

\title{State transfer in static and dynamic spin chains with disorder}

\author{David Petrosyan}
\author{Georgios M. Nikolopoulos}
\author{P. Lambropoulos}
\affiliation{Institute of Electronic Structure and Laser, 
Foundation for Research and Technology - Hellas, 
71110 Heraklion, Crete, Greece}

\date{\today}

\begin{abstract}
We examine the speed and fidelity of several protocols for state 
or single excitation transfer in finite spin chains subject to diagonal 
and off-diagonal disorder. We find that, for a given chain length 
and maximal achievable inter-spin exchange ($XY$) coupling strength, 
the optimal static spin-coupling protocol, implementing the fastest
state transfer between the two ends of the chain, is more susceptible 
to off-diagonal ($XY$ coupling) disorder, as compared to a much slower 
but robust adiabatic transfer protocol with time-dependent coupling 
strengths.
\end{abstract}


\maketitle

\section{Introduction}

Faithful transfer of quantum states between physical qubits 
of an integrated quantum register is one of the important 
prerequisites for scalable quantum computation.
Typically, qubit-qubit interactions are short range and 
implementing quantum logic gates between qubits located at 
distant sub-registers would involve interconnecting them via 
quantum channels, or wires \cite{BoseRev}, which may consist of 
arrays of coupled quantum dots \cite{weNPL,wePL} or superconducting 
qubits \cite{Romito,Strauch,Tsomokos}, atoms in optical 
lattices \cite{SChOL,MVDPepl}, 
or other realizations of spin chains. 

Quantum channels of permanently coupled spins would require no 
dynamical manipulations during the state transfer, but might 
be susceptible to noise and imperfections. Conversely, dynamically
manipulated networks can be more robust with respect to certain
kinds of disorder, but are more involved requiring time-dependent 
external control. Here we re-consider critically several 
protocols for achieving efficient and dependable---ideally perfect---state 
transfer in disordered spin chains subject to physically constrained 
maximal achievable inter-spin coupling rate. The present work is an 
extension of our earlier studies \cite{weNPL,wePL} on perfect states 
transfer to more realistic scenarios with the aim of quantifying and 
neutralizing the influence of static (or slowly changing) noise 
inevitably present in any imperfect physical realization of the 
spin chain resulting in diagonal and off-diagonal disorder.

After outlining the model, we examine the speed and reliability 
of several state transfer protocols first for ideal and then for 
noisy spin chains, followed by conclusions.

\section{The Model}

The Hamiltonian for a spin chain of length $N$ has a general form \cite{Sachdev}
\be
H = \hlf \sum_{j=1}^{N} h_j \hs_j^z 
- \hlf \sum_{j=1}^{N-1} J_j ( \hs_j^x \hs_{j+1}^x + \hs_j^y \hs_{j+1}^y 
+ \De \hs_j^z \hs_{j+1}^z), \label{HamSpCh} 
\ee
where $\hs_j^{x,y,z}$ are the Pauli spin operators at position $j$,
$h_j$ determines the energy separation between the spin-up and 
spin-down states playing the role of the local ``magnetic field'', 
and $J_j$ is the nearest-neighbor spin-spin interaction which
can be static or time-dependent. From now on we set the 
anisotropy parameter $\De=0$; Eq.~(\ref{HamSpCh}) reduces then
to the Hamiltonian of the $XX$ model, which is isomorphic to the 
Hubbard Hamiltonian for spinless fermions or hard-core bosons \cite{Sachdev},
\be
H = \sum_{j=1}^{N} h_j \ha^{\dag}_j \ha_j  
- \sum_{j=1}^{N-1} J_j ( \ha^{\dag}_j \ha_{j+1} + \ha^{\dag}_{j+1} \ha_j), 
\label{HamHub} 
\ee
where $\ha^{\dag}_j$ ($\ha_j$) is the particle creation (annihilation) 
operator at site $j$ with energy $h_j$ and $J_j$ now plays the role
of tunnel couping between adjacent sites $j$ and  $j+1$. 

Our objective here is to transfer an arbitrary single qubit state 
$\ket{\psi} = \alpha \ket{0} + \beta \ket{1}$ between the two ends 
of the spin chain. To that end, we assume that all the spins can 
be prepared in the ``ground'' state $\ket{\ds}_j \equiv \ket{0}_j$ 
and at a certain initial time $t_{\mathrm{in}} =0$ the first site of the 
chain is initialized to $\ket{\psi}_1$. Ideal transfer would imply that 
at a well-defined final time $t_{\mathrm{out}}$ the last site of the 
chain is in state $\ket{\psi}_N$, up to a certain relative phase
factor between the amplitudes of states $\ket{0}_N$ and $\ket{1}_N$
(see below).
 
Since the Hamiltonian (\ref{HamSpCh}) [or (\ref{HamHub})] preserves 
the number of spin- [or particle-] excitations, we need to consider 
only the zero $\ket{\bm{0}} \equiv \prod_{j=1}^N \ket{0}_j$ and 
single excitation $\ket{\bm{j}} \equiv \hs_j^+ \ket{\bm{0}}$ 
[$\ha^{\dag}_j\ket{\bm{0}}$] subspace of the total Hilbert space.
Then the system initially in state 
$\ket{\Psi_{\mathrm{in}}} = \alpha \ket{\bm{0}} + \beta \ket{\bm{1}}$
evolves in time as 
$\ket{\Psi(t)} = U(t) \ket{\Psi_{\mathrm{in}}} = \alpha \ket{\bm{0}} 
+ \beta \sum_{j=1}^N A_j(t) \ket{\bm{j}}$, 
where $U(t) = \mathcal{T} \exp \big[\frac{1}{i \hbar} \int_{0}^t \!\!  
H(t') dt' \big]$ is the (time-ordered, $\mathcal{T}$) evolution operator.
Apparently, only the states in the single 
excitation sub-space $\{\ket{\bm{j}}\}$ evolve in time with the 
corresponding amplitudes $A_j(t) \equiv \bra{\bm{j}} U(t) \ket{\bm{1}}$,
while the vacuum (or ground) state $\ket{\bm{0}}$ remains unchanged. 
Thus perfect state transfer would be achieved for the amplitude 
$|A_N(t_{\mathrm{out}})| =1$, provided its phase $\phi = \arg(A_N)$ 
is fixed and known, $\phi=\phi_0$, and therefore can be amended. 

We may quantify the performance of the scheme by the transfer fidelity 
$F_{\psi} = \bra{\psi} \rho_N \ket{\psi}$ where
$\rho_N \equiv \mathrm{Tr}_{\not N} (\ket{\Psi} \bra{\Psi}) = 
(1- |\beta|^2 |A_N|^2) \ket{0}\bra{0} + |\beta|^2 |A_N|^2 \ket{1}\bra{1} 
+ \alpha \beta^* A_N^* \ket{0}\bra{1} + \alpha^* \beta A_N \ket{1}\bra{0}$ 
is the reduced density operator for the $N$th site of the chain \cite{Bose}. 
We then have $F_{\psi} = |\alpha|^2 + |\beta|^2 (1 - 2 |\alpha|^2) |A_N|^2
+ 2 |\alpha|^2 |\beta|^2 |A_N| \cos(\phi)$,
while the mean transfer fidelity $F$, obtained by averaging $F_{\psi}$ 
over all possible $\ket{\psi}$ and after compensating for $\phi_0$, is 
given by \cite{Bose} 
\be
F = \frac{1}{2} + \frac{|A_N|^2}{6} +  \frac{|A_N| \cos(\phi-\phi_0)}{3}. 
\label{Fav}
\ee
Thus, for the amplitude $|A_N|=1$ but completely random phase $\phi$, 
the fidelity is equal to the classical value of $F = 2/3$, while
for $|A_N|=0$ we have $F = 1/2$ corresponding to a random guess of 
the qubit state $\ket{0}$ or $\ket{1}$.

\section{State transfer protocols}

The state or excitation transfer in a spin chain described by Hamiltonian
(\ref{HamSpCh}) [or (\ref{HamHub})] is mediated by the nearest-neighbor 
couplings $J_j$. Clearly, in any practical realization of the spin 
chain there will be some upper limit for achievable coupling strength, 
$J_{\mathrm{max}} \equiv \max\{ J_j \}$, determined by physical
of technological constraints. On a fundamental level, this follows 
from the fact that the energy of the system is bounded, which, in turn, 
limits the speed of the state transfer, 
$t_{\mathrm{out}} \gtrsim N/J_{\mathrm{max}}$ \cite{Yung}.  

%%%%%%%%%%%%%%%%%%%%%%%
\begin{figure*}[ht]
\centerline{\includegraphics[width=0.65\textwidth]{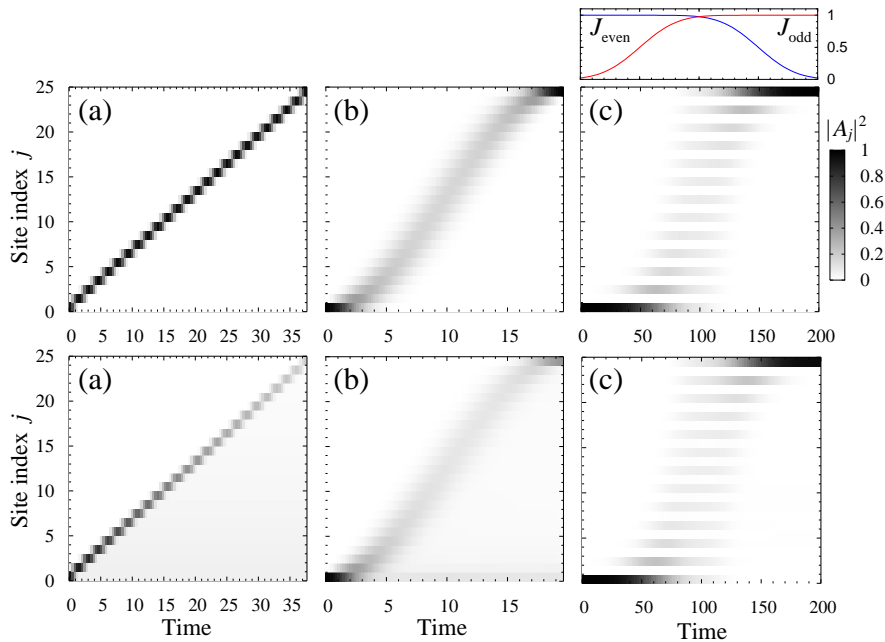}}
\caption{Dynamics of single excitation transfer in a spin chain 
of length $N=25$ for (a) sequential \textsc{swap},
(b) spin-coupling, and (c) adiabatic protocols.
Top panel corresponds to noiseless spin chains, $\sigma_h = \sigma_J = 0$, 
yielding complete excitation transfer $|A_N(t_{\mathrm{out}})|^2 = 1$ 
for all cases (a), (b) and (c). 
The graph above (c) shows the time dependence of couplings 
$J_{\mathrm{even}}$ and $J_{\mathrm{odd}}$ normalized to $J_{\mathrm{max}}$
[cf. Eq.~(\ref{Jodev})].
Bottom panel illustrates the results for noisy chains with 
$\sigma_h = \sigma_J = 0.15 J_{\mathrm{max}}$ averaged over 1000 
independent realizations, leading to 
$\expv{|A_N(t_{\mathrm{out}})|^2} \simeq 0.2$, $0.42$ and $0.96$ 
for (a), (b) and (c), respectively. 
Time is measured in units of $J_{\mathrm{max}}^{-1}$ and the 
evolution terminates at the corresponding $t_{\mathrm{out}}$.}
\label{fig:poptr}
\end{figure*}
%%%%%%%%%%%%%%%%%%%%%%%

\subsection{Noiseless spin chains}

Let us first recall the key facts pertaining to an idealized spin chain 
with no disorder. We assume uniform on-site energies 
$h_j := 0 \; \forall \; j \in [1,N]$, while the individual couplings $J_j$
can be freely controlled, subject to the constraint $J_j \leq J_{\mathrm{max}}$. 

\paragraph*{(a)}
Perhaps conceptually the most straightforward approach to the state transfer 
between the two ends of the chain is to apply a sequence of \textsc{swap} 
operations implemented by $\pi$-pulses between the pairs of neighboring 
sites. To that end, with all the couplings $J_j$ set initially to zero, we
switch on $J_1$ for time $t_1 = \pi/(2J_1)$, then $J_2$ for time 
$t_2 = \pi/(2J_2)$, etc. till reaching the $N$th site. At the end 
of each step, the corresponding state amplitude is 
$A_j(t_{j-1}) = -i \sin(J_{j-1}t_{j-1}) \, A_{j-1}(t_{j-2})= (-i)^{j-1}$ 
for $j=2,\ldots, N$. If all the couplings' strengths can be pulsed 
to the maximal possible $J_{\mathrm{max}}$, and there are $N-1$ steps, 
the total transfer time is $t_{\mathrm{out}} = (N-1)\pi/(2J_{\mathrm{max}}) 
\simeq (\pi/2) (N /J_{\mathrm{max}})$ ($N \gg 1$)
with the final state amplitude $A_N(t_{\mathrm{out}}) = (-i)^{N-1}$,
i.e., $|A_N(t_{\mathrm{out}})| =1$ and $\phi_0 = (-\pi/2)(N-1) \pmod{2\pi}$ .

\paragraph*{(b)}
We next consider a spin chain with static couplings $J_j$ arranged in
an appropriate way facilitating the perfect state (or excitation) transfer.
By ``static'' we mean that during the transfer the coupling strengths
are fixed, but to initiate (at time $t_{\mathrm{in}}$) and to terminate 
(at time $t_{\mathrm{out}}$) the transfer process at least $J_1$ and 
$J_{N-1}$ should be quickly switched on and off, respectively. 
(Alternatively the state initialization of the first site at $t_{\mathrm{in}}$
and state retrieval from the last site at $t_{\mathrm{out}}$ should be 
accomplished very fast, on a time-scale short compared to $J_{1,N-1}^{-1}$.) 
Among the many \cite{BoseRev}---in fact, infinitely many 
\cite{Kostak}---possible static protocols for perfect state transfer, 
we focus here on the one proposed in \cite{weNPL,Ekert}, and much earlier 
\cite{CookShore} and in a different context (that of population transfer in 
laser-driven multilevel atomic or molecular systems\cite{ShoreBook}), which
was shown to be the optimal one \cite{Yung} in terms of the transfer time. 
In this so-called spin-coupling protocol, the coupling constants are 
arranged according to $J_j = J_0 \sqrt{(N-j)j}$, which makes the system 
formally analogous  to a spin-$\Js$ in a magnetic field \cite{spinJ}. 
This leads to the equidistant energy spectrum $\la_k = 2J_0 k - J_0 (N+1)$ 
with $k = 1,2,\ldots, N$, and consequently perfectly periodic oscillations 
of the single excitation between the two ends of the chain, according to 
\[
A_j(t) = \left(\begin{array}{c}
N-1 \\ j-1 \end{array} \right)^{1/2} 
[-i \sin{(J_0 t)}]^{(j-1)} \cos{(J_0 t)}^{(N-j)} .
\]
Thus, at time $t_{\mathrm{out}} = \pi/(2 J_0)$ the amplitude of
the final state is $A_N(t) = [-i \sin(J_0 t_{\mathrm{out}})]^{N-1} 
= (-i)^{N-1}$. Note that the strongest coupling is in the center of the chain:
at $j = N/2$ for $N$ even, $J_{N/2} = \hlf J_0 N \equiv J_{\mathrm{max}}$;
or at $j = (N \pm 1) /2$ for $N$ odd, $J_{(N\pm 1)/2} = \hlf J_0 \sqrt{N^2-1}
\simeq J_{\mathrm{max}}$ ($N \gg 1$). Hence, the transfer time expressed 
through $J_{\mathrm{max}}$ is given by $t_{\mathrm{out}} 
= (\pi/4) (N /J_{\mathrm{max}})$, which is twice shorter than that 
for the sequential \textsc{swap} protocol. 

\paragraph*{(c)}
The last protocol that we consider here is the adiabatic state or excitation 
transfer between the two ends of the spin chain using slowly varying 
couplings $J_j$ \cite{wePL,GCHH}. This is analogous to the stimulated Raman 
adiabatic passage (STIRAP) techniques \cite{stirap-rev} extended to 
multilevel atomic or molecular systems \cite{stirapNsqLs}. 
Assume that $N$ is odd and the individual couplings $J_j$ can be 
selectively and independently manipulated. In the single excitation subspace, 
the Hamiltonian (\ref{HamSpCh}) [or (\ref{HamHub})] has an eigenstate 
\begin{align}
\ket{\Psi^{(0)}} = &\frac{1}{\sqrt{\mathcal{N}_0}} 
[J_2 J_4 \ldots J_{N-1} \ket{\bm{1}} +
(-1) J_1 J_4 \ldots J_{N-1} \ket{\bm{3}} + 
\nonumber \\ & \quad \ldots 
+ (-1)^{\Js} J_1 J_3 \ldots J_{N-2} \ket{\bm{N}}],  \label{CPTNl} \\
& \Js \equiv \hlf (N-1) , \nonumber
\end{align}
with eigenvalue $\la^{(0)}=0$, which is conventionally called coherent 
population trapping (or dark) state \cite{stirap-rev,stirapNsqLs}. 
Thus the amplitude of initial state $A_1$ is proportional 
to the product of all the even-numbered couplings, while the amplitude 
of final state $A_N$ is given by the product of all the 
odd-numbered couplings, divided by the normalization parameter 
$\mathcal{N}_0 = (J_2 J_4 \ldots J_{N-1})^2 + \ldots 
+ (J_1 J_3 \ldots J_{N-2})^2$. Therefore, if all the even-numbered
couplings are switched on first, the zero-energy state (\ref{CPTNl}) 
would coincide with the initial state $\ket{\bm{1}}$.
This is then followed by adiabatically switching-on of all the odd-numbered 
couplings, while the even-numbered couplings are switched-off, which will 
result in state (\ref{CPTNl}) to rotate towards the final state 
$\ket{\bm{N}}$. Assuming that these two families of couplings
are described by common shape functions, 
$J_2, J_4, \ldots , J_{N-1} = J_{\mathrm{even}} (t)$ 
and $J_1, J_3, \ldots , J_{N-2} = J_{\mathrm{odd}}(t)$, 
the amplitudes of the initial and the final states are given by 
\[
A_1(t) = \frac{[J_{\mathrm{even}}(t)]^{\Js}}{\sqrt{\mathcal{N}_0(t)}} , \quad  
A_N(t) = (-1)^{\Js} \frac{[J_{\mathrm{odd}}(t)]^{\Js}}{\sqrt{\mathcal{N}_0(t)}} ,
\]
with $\mathcal{N}_0(t) = \sum_{n=0}^{\Js} 
[J_{\mathrm{odd}}(t)]^{2n} [J_{\mathrm{even}}(t)]^{2(\Js - n)}$. 
Thus, complete state or excitation transfer between the two ends of 
the chain can be achieved by applying first the $J_{\mathrm{even}}$ couplings
and then the $J_{\mathrm{odd}}$ couplings, the two sets of couplings
partially overlapping in time. At time $t_{\mathrm{out}}$, when
$J_{\mathrm{odd}}(t_{\mathrm{out}}) \gg J_{\mathrm{even}}(t_{\mathrm{out}}) \simeq 0$,
the amplitude of the final state is $A_N(t_{\mathrm{out}}) = (-1)^{\Js}$,
i.e., $|A_N(t_{\mathrm{out}})| =1$ and $\phi_0 = (-\pi)(N-1)/2 \pmod{2\pi}$. 
Of course the adiabatic following of the zero-energy eigenstate (\ref{CPTNl}) 
holds true if, during the transfer process, the non-adiabatic transitions out
of $\ket{\Psi^{(0)}}$ are negligible, which requires that the rate of change 
of the coupling strengths be small compared to energy separation between 
$\ket{\Psi^{(0)}}$ and all the other eigenstates. 
We can estimate the energy separation between the eigenstates 
in the vicinity of maximal overlap between the even and odd couplings, 
$J_{\mathrm{even}} \simeq J_{\mathrm{odd}} = J$.
The energy spectrum of the chain with homogeneous coupling, 
$J_j= J \; \forall \; j \in [1,N]$, is $\la_k = - 2 J \cos [k \pi/(N+1)]$ 
\cite{weNPL,wePL}. The eigenstate with zero energy $\la^{(0)}$ is 
the one with $k = (N+1)/2 \equiv k_0$, and the nearest eigenstates 
with indices $k = k_0 \pm 1$ have energies 
$\la_{k_0 \pm 1} = \pm 2 J \sin [\pi/(N+1)] \simeq \pm 2 J \pi /N$ ($N \gg 1$).
With $J \lesssim J_{\mathrm{max}}$, the excitation transfer time 
$t_{\mathrm{out}}$, being roughly equal to the couplings' switching time,
should then satisfy the condition $t_{\mathrm{out}} \gg N/(2 \pi J_{\mathrm{max}})$.

%%%%%%%%%%FIGURE%%%%%%%%%%%%
%%%%%%%%%%%%%%%%%%%%%%%%%%%%

To summarize the results for noiseless spin chains, 
the transfer time $t_{\mathrm{out}}$ for all three protocols 
scales with the number of sites $N$ and the maximal inter-site coupling 
$J_{\mathrm{max}}$  as $(N/J_{\mathrm{max}})$. The fastest is the 
spin-coupling protocol with $t_{\mathrm{out}} = (\pi/4) (N /J_{\mathrm{max}})$.
It is followed by the sequential \textsc{swap} protocol, for which
$t_{\mathrm{out}} \simeq (\pi/2) (N /J_{\mathrm{max}})$.  
Finally the slowest is the adiabatic protocol 
$t_{\mathrm{out}} \simeq C \, (N/J_{\mathrm{max}})$, with $C \simeq 8$ 
being a safe estimate \cite{Ccomm} for smooth coupling functions
that we use:
\be
J_{\substack{ \mathrm{odd} \\ \mathrm{even}}}(t) = J_{\mathrm{max}} \,
\frac{1}{2} \left[ 1 
\pm \erf \left( \frac{t - \frac{1}{2}t_{\mathrm{out}} \pm 2 \sigma_t}
{\sqrt{2} \sigma_t} \right) \right] , \label{Jodev}
\ee
with $\sigma_t = \frac{1}{8}t_{\mathrm{out}}$.
Note that the phase of the final state amplitude for all
three protocols is given by $\phi_0 = (-\pi/2)(N-1) \pmod{2\pi}$, which
should be compensated for after the transfer.  
The single excitation transfer for all three protocols
is illustrated in the top panel of Fig.~\ref{fig:poptr}.

\subsection{Disordered chains}

%%%%%%%%%%%%%%%%%%%%%%%
\begin{figure*}[ht]
\centerline{\includegraphics[width=0.8\textwidth]{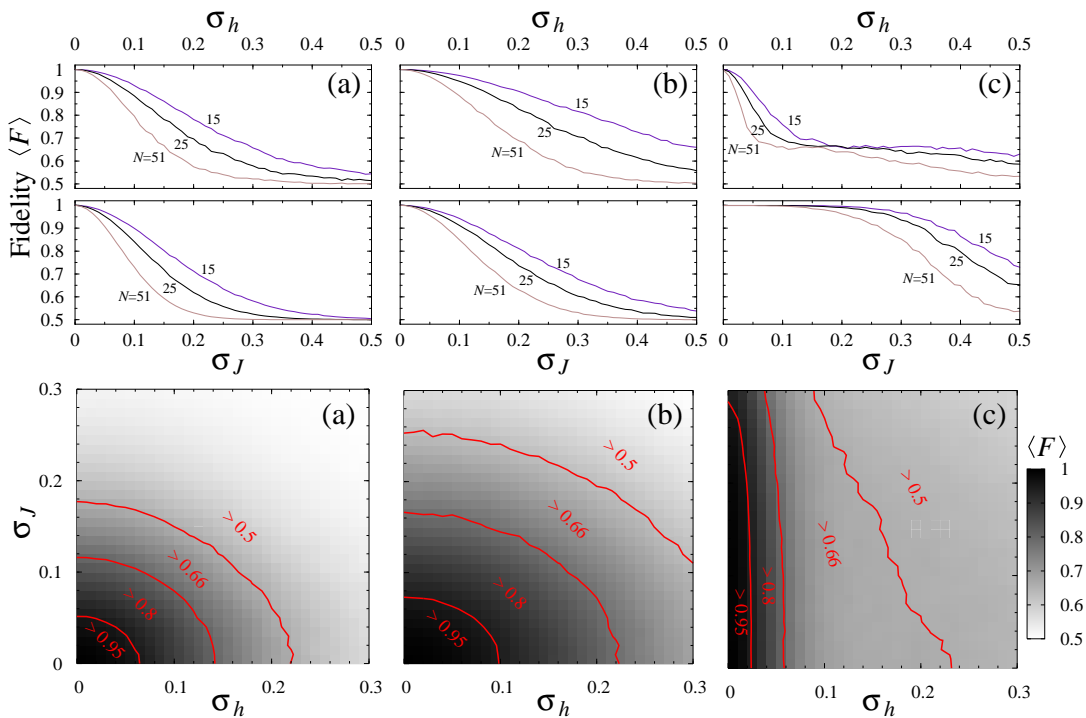}}
\caption{Averaged (over 1000 realizations) fidelity $\expv{F}$ 
in noisy spin chains for (a) sequential \textsc{swap}, 
(b) spin-coupling, and (c) adiabatic protocols.
Top panel shows the dependence of $\expv{F}$ on the diagonal 
disorder $\sigma_h$ with $\sigma_J =0$ (upper plots), and
on the off-diagonal disorder $\sigma_J$ with $\sigma_h =0$ (lower plots),
for the chains of lengths $N=15,25,51$. 
Bottom panel shows $\expv{F}$ versus both $\sigma_h$ with $\sigma_J$
in spin chains with $N=25$.}
\label{fig:fidels}
\end{figure*}
%%%%%%%%%%%%%%%%%%%%%%%%

Employing numerical simulations, we now examine robustness of the above 
described state transfer protocols in spin chains with varying degree 
of disorder. We note that for the spin-coupling scheme, related analysis
has been performed in \cite{DeChiara}. 

The physical origin of disorder may be two-fold: 
(i) fabrication imperfections of the particular system 
realizing the spin chain, and 
(ii) noise of the external controls, which is assumed to vary slowly 
enough on the time-scale of state transfer $t_{\mathrm{out}}$, 
as is typically the case in most experimental situations 
pertaining to coupled quantum dots \cite{QDsRev}, superconducting
qubits \cite{SCQsRev} or atoms \cite{OptLatRev}. 
We shall distinguish diagonal and off-diagonal disorder. 
The diagonal disorder corresponds to random on-site energies, 
or equivalently the local magnetic fields $h_j$, normally 
distributed around $\expv{h_j} =0$ with variance $\sigma_h^2$
(without loss of generality, we assume that the energies 
of the first and the last sites of the chain are exempt 
from disorder, $h_1=h_N =0$, since otherwise the state 
$\ket{\psi}$ would dephase even before and after the transfer). 
The off-diagonal disorder introduces randomness in the
inter-site coupling strengths $J_j \to J_j (1 + \de J_j)$
where $\de J_j$ are normally distributed around $\expv{\de J_j} = 0$ 
with variance $\sigma_J^2$.  
Consistently with the above description, we will treat the disorder 
as static during each realization of the numerical experiment for 
the particular protocol, but completely uncorrelated between different
realizations. The results presented below are obtained by averaging 
over many (typically 1000) independent realizations.

%%%%%%%%%%FIGURE%%%%%%%%%%%%
%%%%%%%%%%%%%%%%%%%%%%%%%%%%

Figure~\ref{fig:poptr} compares the single excitation transfer for 
the three protocols (a), (b) and (c) in ideal and disordered spin chains 
of length $N=25$. In the noiseless chain we have perfect transfer 
$|A_N(t_{\mathrm{out}})|^2 = 1$, while in the presence of diagonal 
and off-diagonal disorder characterized by standard deviations 
$\sigma_h = \sigma_J = 0.15 J_{\mathrm{max}}$, the averaged transfer 
probabilities are reduced to $\expv{|A_N(t_{\mathrm{out}})|^2} \simeq 0.2$, 
$0.42$ and $0.96$ for the cases of (a), (b) and (c), respectively. 
Thus, among the three transfer protocols, the sequential \textsc{swap} 
scheme is the most susceptible to noise, especially to the off-diagonal 
disorder which leads to deviations of the subsequent pulse areas from 
the required value of $\pi$; in this particular example, the off-diagonal 
disorder alone is responsible for at least 70\% reduction of the 
transfer probability. The spin-coupling scheme is somewhat more
robust with respect to noise, with both diagonal and off-diagonal 
disorder comparably contributing to the reduction of the transfer 
probability by about 20\% and 40\%, respectively. Finally, the 
adiabatic transfer scheme is very tolerant to noise, as far as the
transfer probability is concerned, but is quite slow; in fact
it can tolerate even more disorder at the expense of slowing it further 
down (equivalent to increasing $C$). 

The probability of excitation transfer alone is not enough to fully
characterize the state transfer, since, e.g., $|A_N(t_{\mathrm{out}})|^2 = 1$ 
but completely random phase $\phi$ amounts to classical information 
transfer only, and the resulting fidelity for quantum state transfer 
is merely $F= 0.66$. We will therefore quantify the performance of 
the system subject to varying level of noise using fidelity $\expv{F}$ 
of Eq.~(\ref{Fav}) averaged over many independent realizations of 
protocols (a), (b) and (c). Figure~\ref{fig:fidels} summarizes 
the results of our numerical simulations for the chains of lengths $N=15$, 
$25$ and $51$. Unsurprisingly, the longer the chain the lower the fidelity of 
the state transfer. We find that, for the same values of diagonal $\sigma_h$
and/or off-diagonal $\sigma_J$ disorder, the sequential \textsc{swap} 
scheme yields lower fidelity than the spin-coupling scheme. Moreover, 
both schemes are somewhat more susceptible to the off-diagonal disorder.
The behavior of the fidelity for the adiabatic transfer scheme 
is, however, profoundly different: it is very robust with respect
to the off-diagonal disorder $\sigma_J$, but much more sensitive to 
the diagonal disorder $\sigma_h$: already for $\sigma_h \gtrsim 0.1$ the 
fidelity $\expv{F} \simeq 0.66$ (but then decreases slowly with increasing
$\sigma_{h}$). This is despite the fact that the transfer probability
$\expv{|A_N(t_{\mathrm{out}})|^2}$ remains above $0.9$ up to 
$\sigma_{h,J} \lesssim 0.28$, i.e., the excitation transfer is very 
efficient up to large values of both diagonal and off-diagonal disorder.
The adiabatic transfer protocol is so sensitive to diagonal disorder 
because it is slow: during the long transfer time $t_{\mathrm{out}}$ 
even little noise in the on-site energies $\sigma_h$
accumulates to large random phase $\phi$ spread over 
$\sigma_{\phi} \sim \sigma_h t_{\mathrm{out}}$.

\section{Conclusions}

We have critically examined the state and excitation transfer in 
disordered spin chains using the sequential \textsc{swap},
spin-coupling and adiabatic transfer protocols. We have found that,
depending on the character of disorder, namely the diagonal disorder 
corresponding to random on-site energies (or magnetic filed) 
or off-diagonal disorder leading to variations in inter-site couplings, 
either the fast spin-coupling protocol or the slow adiabatic transfer 
protocol is more suitable for high-fidelity transfer of quantum 
states between the two ends of the spin chain. 

Reliable quantum channels, based on, e.g., spin chains, are 
indispensable for achieving scalable and efficient quantum 
information processing in solid-state systems with fixed qubit 
positions and finite-range inter-qubit interactions. 
Our results therefore have important implications for attaining 
scalability in such systems.

\begin{acknowledgments}
This work was supported by the EC Marie Curie Research-Training 
Network EMALI (MRTN-CT-2006-035369). 
\end{acknowledgments}


\begin{thebibliography}{99}
\bibitem{spinJ}
Recall from the theory of angular momentum that a spin-$\Js$ 
particle is a constant magnetic field exhibits 
non-dispersive (Larmor) precession about the field direction. 
With the quantization direction along an axis perpendicular to the 
magnetic field direction, the matrix elements for the transitions 
$\ket{\Js,m} \lra \ket{\Js,m+1}$ between the neighboring magnetic  
sub-states ($m = - \Js, \ldots, \Js$) are proportional to 
$\sqrt{(\Js-m)(\Js+m+1)}$. 
Setting formally $N = 2\Js+1$ and $j = \Js + m +1$, leads 
to the coupling constants $J_j \propto \sqrt{(N-j)j}$, as in the text.
\bibitem{Ccomm}
We note that the smallest value of $C$ leading to good adiabatic
following is quite sensitive to the choice of coupling functions 
$J_{\mathrm{odd}}(t)$ and $J_{\mathrm{even}}(t)$ and their overlap.
(For the particular choice of the coupling functions, even for $C=5$ 
we attain good adiabatic following with excitation transfer 
$|A_N(t_{\mathrm{out}})|^2 \gtrsim 0.98$.) Moreover, with increasing 
the chain length $N$, $\min (C)$ slowly grows with $N$, because the 
latter determines not only the energy separation $2 J \pi /N$ of 
the nearest eigenstate from the zero-energy eigenstate, but also 
the number of further detuned eigenstates to which non-adiabatic 
transitions could occur, albeit with even lesser probabilities.
\end{thebibliography}
\end{document}